\documentclass{aa}  

\usepackage{graphicx}
\usepackage{txfonts}
\begin{document} 

\title{Outstanding X-ray emission from the stellar radio pulsar CU~Virginis}

\author{J. Robrade\inst{1}
\and L.M. Oskinova\inst{2,3}
\and J.H.M.M. Schmitt\inst{1}
\and P. Leto\inst{4}
\and C. Trigilio\inst{4}
}
\institute{Hamburger Sternwarte, University of Hamburg, Gojenbergsweg 112, D-21029 Hamburg, Germany
\and Institute for Physics and Astronomy, University Potsdam, D-14476 Potsdam, Germany
\and Kazan Federal University, Kremlevskaya Str 18, Kazan, Russia
\and INAF - Osservatorio Astrofisico di Catania, Via S. Sofia 78, I-95123 Catania, Italy\\
\email{jrobrade@hs.uni-hamburg.de}
}
\date{Received May 24, 2018; accepted xxx}

\abstract
{Among the intermediate mass, magnetic chemically peculiar (MCP) stars, CU~Vir is one of the most intriguing objects. Its 100\,\% circularly polarized beams of radio emission sweep the Earth as the star rotates, thus making this strongly magnetic star the prototype of a class of non degenerate stellar radio pulsars. While CU~Vir is well studied in radio, its high-energy properties are not known. Yet, X-ray emission is expected from stellar magnetospheres and confined stellar winds.}
{Using X-ray data we aim to test CU Vir for intrinsic X-ray emission and investigate mechanisms responsible for its generation.}
{
We present X-ray observations performed with {\it XMM-Newton} and {\it Chandra} and study obtained X-ray images, light curves and spectra. Basic X-ray properties are derived from spectral modelling and are compared with model predictions. In this context we investigate potential thermal and non-thermal X-ray emission scenarios.
}
{We detect an X-ray source at the position of CU~Vir. With $L_{\rm X} \approx 3 \times 10^{28}$~erg\,s$^{-1}$ it is moderately X-ray bright, but the spectrum is extremely hard compared to other Ap stars. Spectral modelling requires multi-component models with predominant hot plasma at temperatures of about $T_{\rm X}= 25$~MK or, alternatively, a nonthermal spectral component. Both types of model provide a virtually equivalent description of the X-ray spectra.
The {\it Chandra} observations was performed six years later than the one by {\it XMM-Newton}, yet the source has similar X-ray flux and spectrum, suggesting a steady and persistent X-ray emission. This is further confirmed by the X-ray light curves that show only mild X-ray variability.
}
{CU~Vir is also at X-ray energies an exceptional star. To explain its full X-ray properties, a generating mechanism beyond standard explanations like the presence of a low-mass companion or magnetically confined wind-shocks is required. Magnetospheric activity might be present or, as proposed for fast rotating strongly magnetic Bp stars, the X-ray emission of CU~Vir is predominantly auroral in nature.
}

\keywords{stars: individual: CU Vir, stars: activity, stars: chemically peculiar, stars: magnetic field --  X-rays: stars}

\maketitle
%

\section{Introduction}

The A0p star CU Vir (HD 124224, HR 5313) is an enigmatic star of the upper part of the main sequence located at a distance of about 79~pc \citep{lee07}. With its $V= 5.0$~mag it is among the prominent, nearby, magnetic, intermediate mass, chemically peculiar stars (MCP star, ApBp star) that shows pronounced photometric and spectroscopic variability \citep[see e.g.][]{bab58}. Furthermore, it is so far a unique
main sequence star that shows regular radio pulses persisting over decades, resembling the radio lighthouse of pulsars and interpreted as auroral radio emission similar to those observed on planets \citep[see e.g.][]{tri00, tri11}. In contrast to the rotationally modulated gyrosynchrotron radio emission commonly observed in MCP stars, the 100\,\% right-handed circularly polarized radio pulses from CU~Vir are explained by the electron cyclotron maser emission (ECME) mechanism. 
The polarization sense indicates that these originate from the northern polar regions of the oblique stellar magnetosphere, where annular rings emit narrow radio beams that sweep over the Earth location twice per stellar rotation. With a period of about $P_{\rm rot} = 0.52$~d, CU Vir is an unusually fast rotator for its class that shows in addition alternating variability of its rotation period over decades \citep{mik11,krt17}.

Beside numerous radio studies, several dedicated multiwavelength campaigns on CU Vir were executed. A detailed study of its variability in optical/UV emission and its spectral energy distribution showed that this variability can be explained by strong spots of elemental over- and under-abundances \citep{krt12}. 
A mapping of CU~Vir's abundance anomalies and magnetic field based on spectropolarimetric observations as well as a reassessment of the stellar parameters was recently performed in \cite{koch14}. According to this study,
CU~Vir is viewed under an inclination of about $i = 46^{\circ}$; its magnetic dipole axis is tilted by about $\beta =79^{\circ}$ to the rotational axis and has a field strength of $B_{d}= 3.8$~kG. 
The derived magnetic maps show a dipolar-like field topology that is non-axisymmetric with large differences between regions of opposite polarity, providing a natural explanation for the north-south asymmetry in the observed radio pulses.

The spectral classification of CU~Vir (B9p\,/\,A0p) as well as its stellar parameters vary a bit in literature, recent values as determined by \cite{koch14} are $M=3.1 M_{\odot}$, $R=2.1 R_{\odot}$  and $T_{\rm eff}=12750$~K. Using $Vsini=145$~km\,s$^{-1}$ this implies with $i = 46^{\circ}$ a rotational speed of about 200~km\,s$^{-1}$. While rotating fast, probably due to its youth, CU~Vir is still nearly spherical and effects of gravitation darkening are moderate with temperature contrast of only a few 100~K.

The existence of intrinsic X-ray emission from late-B and early-A stars has been debated, as
'normal' late-B\,/\,early-A stars are expected to be virtually X-ray dark, since these stars neither drive magnetic activity nor strong stellar winds. Often, low-mass coronal companions are suspected and identified as the true X-ray source. However, intrinsic X-ray emission from magnetically confined wind shocks (MCWS) could be also expected in the Ap/Bp stars. In this model, originally proposed to explain the X-ray emission of the A0p star IQ~Aur \citep{bab97}, the stellar wind from both hemispheres is channelled by the magnetic field and collides in the vicinity of the equatorial plane, leading to strong shocks and thereby plasma heating to X-ray temperatures. Advanced MCWS models using magneto-hydrodynamic simulations (MHD) or rigid rotating magnetospheres models (RRM) are now 'standard' models and have been used to interpret the observed X-ray properties of magnetic massive stars \citep[e.g.][]{dou14}.

A 3-D model able to simulate the the incoherent gyrosynchrotron radio emission from a typical rapidly rotating magnetosphere of a hot magnetic star has been previously developed \citep{tri04}.
Using this model, the multiwavelength radio light curves of CU~Vir have been simulated, constraining the magnetospheric physical conditions \citep{leto06}. Recently, the same simulation approach has been successfully applied also to the cases of the fast rotating and strongly magnetic B2Vp stars HR7355 and HR5907 \citep{leto17,leto18}.
These two stars show evidences of non-thermal X-ray emission.
By using the 3-D model, computed to simulate the gyro-synchrotron radio emission of these two stars, it was shown, that in addition to thermal plasma heated by the shocked magnetically confined wind streams, a nonthermal auroral X-ray radiation is also expected. Assuming a common framework able to explain the radio emission features of the hot magnetic stars, it is expected that similar auroral X-ray emission shall be observable also from CU~Vir.

The star CU Vir was observed by {\it XMM-Newton} in 2011 and detected in X-rays \citep{rob16}, but the combination of positional offset and fuzziness of PSF made a confirmation of the detection desirable. For this purpose a {\it Chandra} ACIS observation was initiated by us and performed in 2017. Here we present a detailed analysis of the available X-ray data from CU Vir and put it into context of current models to explain the X-ray emission in magnetic intermediate mass stars.

\section{Observations and data analysis}
\label{obs}

The target CU Vir was observed with {\it XMM-Newton} for about 30~ks in 2011 (ToO, PI: Schartel) and we use data from the EPIC camera, consisting of the pn and two MOS detectors. All detectors were operated with the thick filter to avoid optical contamination. The OM was blocked and the RGS produced no useful data due to the low source flux.
Motivated by X-ray excess photons present in the {\it XMM-Newton} data, we initiated a re-observation with {\it Chandra} (PI: Robrade) that was performed in 2017 with the ACIS-S detector in the 1/4 array configuration.
While {\it XMM-Newton} has a higher sensitivity, {\it Chandra} has a much better spatial resolution and is less prone to background events. The two observations complement each other very well. 

A detailed description of the instruments can be found on the respective mission websites ({\it https://www.cosmos.esa.int/web/xmm-newton; http://cxc.harvard.edu/index.html}), the used observations are summarized in Table~\ref{obslog}.

\begin{table}[t]
\begin{center}
\caption{\label{obslog} X-ray observation log for CU Vir.}
\begin{tabular}{lrrr}\hline\\[-3mm]
Mission &  Obs. ID. &Start-Date & Texp [ks]\\\hline\\[-3mm]
{\it XMM-Newton}  & 0677980501 & 2011-07-17 & 27 \\
{\it Chandra} &18925 &2017-05-13 & 29\\\hline
\end{tabular}
\end{center}
\end{table}

\begin{figure}[t]
\includegraphics[width=89mm]{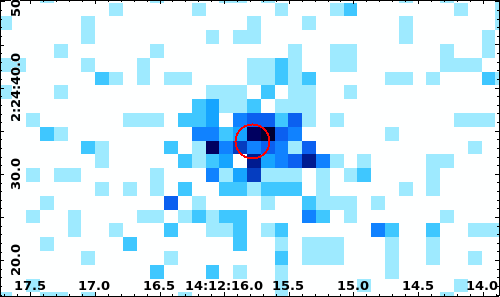}

\vspace*{1mm}
\includegraphics[width=89mm]{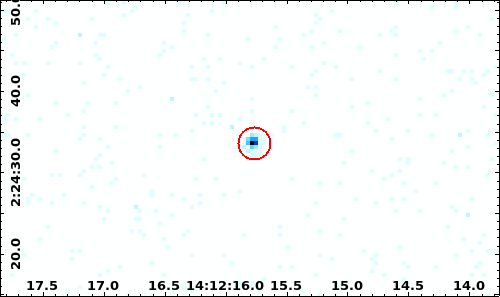}
\caption{\label{ximg} X-ray images of CU Vir; {\it XMM-Newton} pn (top), {\it Chandra} ACIS-S (bottom). The red circle (2\arcsec~radius) denotes the resp. optical position.}
\end{figure}

The {\it XMM-Newton} data were reprocessed with the {\it XMM-Newton} Science Analysis System software SAS~16.1 \citep{sas} and standard SAS tools were used to produce images, light curves, and spectra.
The {\it Chandra} data processing used the CIAO~4.9 software package \citep{ciao} and again standard tools were used to produce images, light curves, and spectra.
Source photons were extracted from circular regions, background was taken from nearby regions. The {\it XMM-Newton} detectors cover the energy range 0.2\,--\,10.0~keV, while ACIS-S covers nominally the 0.3\,--\,10.0~keV range, but its effective area below 0.5~keV is very small. 

Spectral analysis was carried out with XSPEC V12.9 \citep{xspec} and we used multi-temperature APEC \citep{apec} with solar abundances as well as APEC+powerlaw models
to fit the X-ray spectra. The used spectral models assume a solar abundance pattern and ignore a potential absorption component as is was found to be consistent with zero. Adding further spectral components or introducing additional free parameters do not improve the models significantly and result in poorly constrained values. For the {\it XMM-Newton} observation the pn and the combined MOS spectrum are modeled simultaneously. All spectra were rebinned to a minimum of 5 counts per bin and model optimization used the 'cstat' algorithm, applicable for Poisson distributed data. This approach allows to treat the data from the different detectors in an identical fashion without loosing much spectral resolution. We tested different binning and modelling approaches and obtained very similar results, indicating the robustness of our general findings. Given errors denote the 90\% confidence range.

\section{Results}
\label{res}
Here we report on the results obtained from the X-ray observations of CU~Vir, subdivided into separate topics.

\subsection{X-ray images and light curves}

\begin{figure}[t]
\includegraphics[width=90mm]{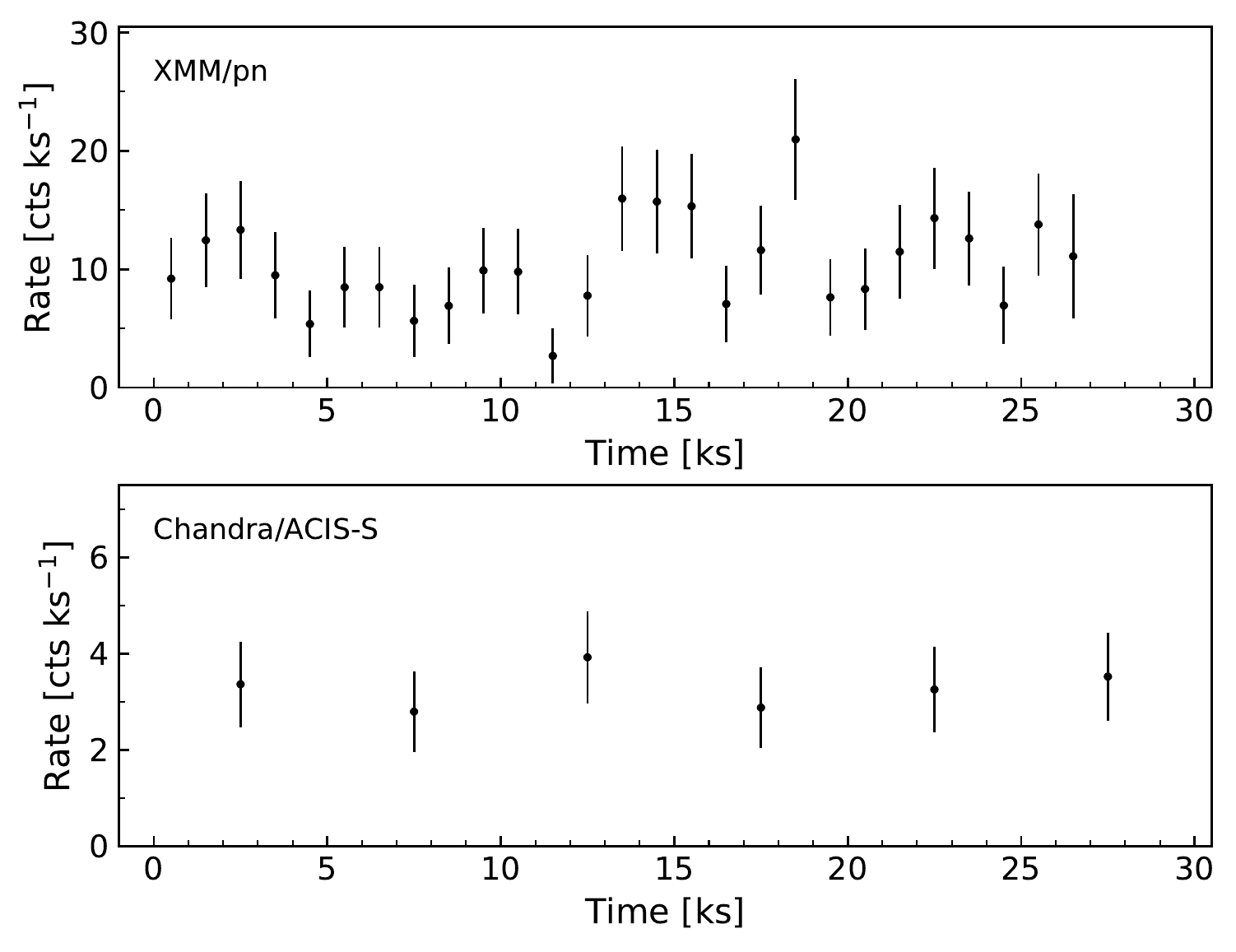}
\caption{\label{xlcs} X-ray light curves of CU Vir; {\it XMM-Newton} pn data with 1~ks binning (top) and {\it Chandra} ACIS-S with 5~ks binning (bottom).}
\end{figure}

A clear photon excess at the expected source position is present in both X-ray images of CU~Vir, that are shown in Fig.~\ref{ximg}. The {\it XMM-Newton} source photon distribution is a bit fuzzy and its photon centroid is about 1.5\arcsec offset from the optical position. While this is not uncommon for {\it XMM-Newton} pn (EEF80/1keV is about 20\arcsec), with EEF80 denoting an encircled energy fraction of 80\,\%, for a clear association of the detected X-ray photons to a single source at the position of CU~Vir a confirmation is desirable. {\it Chandra} is well suited for this exercise (EEF80/1keV is about 0.5\arcsec) and the ACIS-S image shows a source that is point like at the spatial resolution of the detector. Its position matches the optical position at about 0.4\arcsec, i.e. a typical value for the absolute positional accuracy (68\% limit is 0.5\arcsec). No other X-ray source is present in its vicinity, therefore we attribute the detected X-ray photons to CU~Vir in both cases. Whether CU~Vir itself is a true single star cannot be determined with X-ray data, but contaminating sources at separations above a few tenths of an arcsecond can be ruled out.

Next, we inspect the X-ray light curves of CU Vir to investigate potential temporal variability of the source. In Fig.~\ref{xlcs} we show the background subtracted light curves over the full energy band of each detector. The low number of detected counts hinders a detailed variability study on short timescales, but strong peaks in X-ray brightness or a pronounced rotational modulation are clearly not present in our data. Overall, a mostly constant as well as a moderately variable source with brightness changes up to a factor of about two are consistent with the data.

CU~Vir has a rotation period of about 0.52~d (45~ks) and each X-ray observations covers a bit more than half a stellar rotation period.
To roughly put the X-ray data into the rotational frame of CU Vir, we use the ephemeris given in \cite{tri11}.
The radio emission peaks around $\phi = 0.35-0.4$ and $\phi = 0.75-0.8$, slightly offset but close to the phases of zero longitudinal magnetic field. The stronger and sharper negative, southern pole is best visible at $\phi \approx 0.6$, the weaker, positive magnetic pole at $\phi \approx 0.1$. 
The {\it XMM-Newton} pn exposure starts at $\phi = 0.12$ and the radio maximum occurs about 12~ks into the X-ray observation. The X-ray flux during this period is rather low and, if any, the X-ray brighter phases coincide with the phases when the magnetic pole is best visible.
In contrast, the {\it Chandra} observation start around $\phi = 0.65$ and covers basically the rotational phase where the stronger, negative magnetic pole dominates at the beginning of the observation.

\subsection{X-ray spectra}

\begin{figure}[t]
\includegraphics[width=90mm]{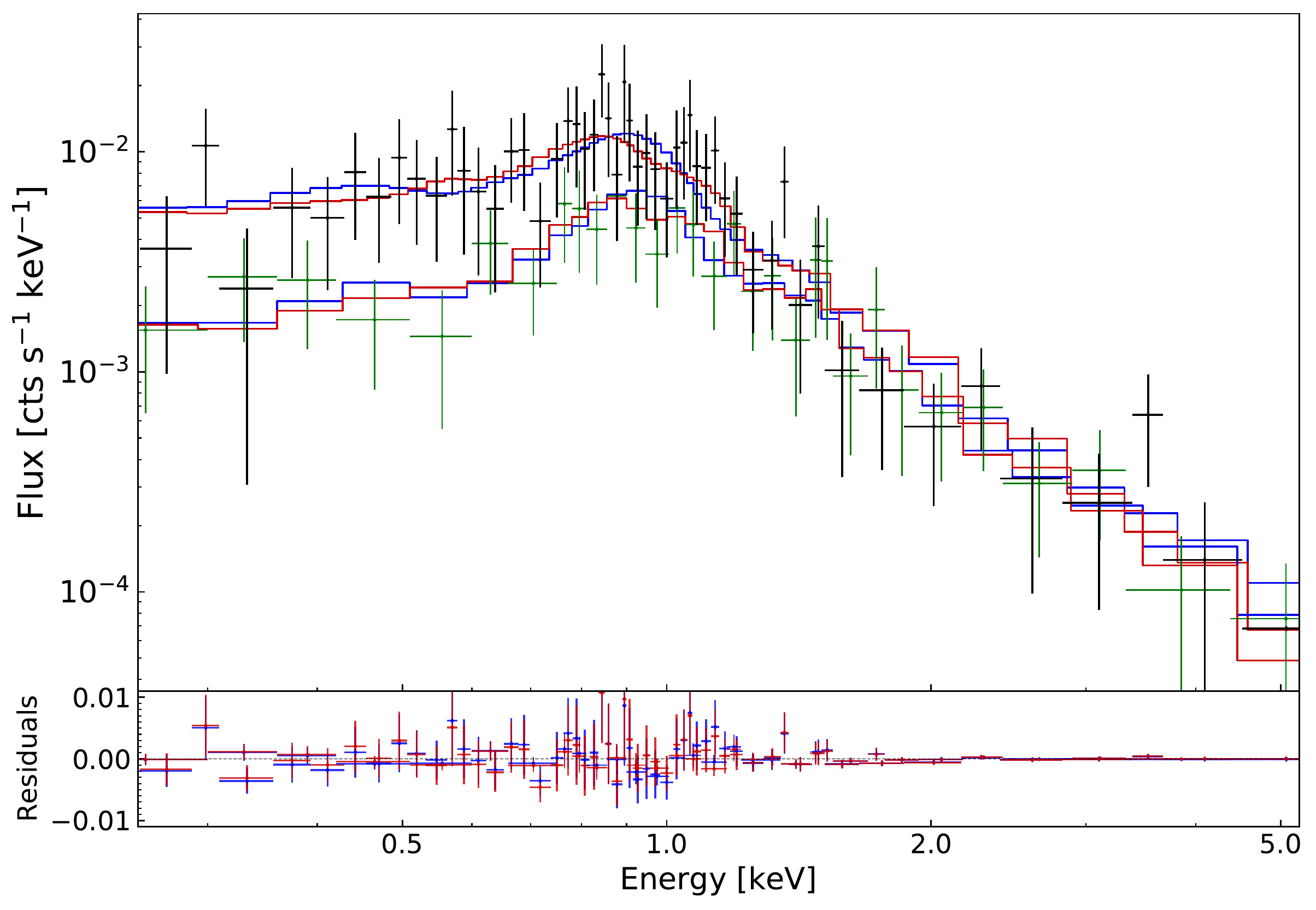}

\includegraphics[width=90mm]{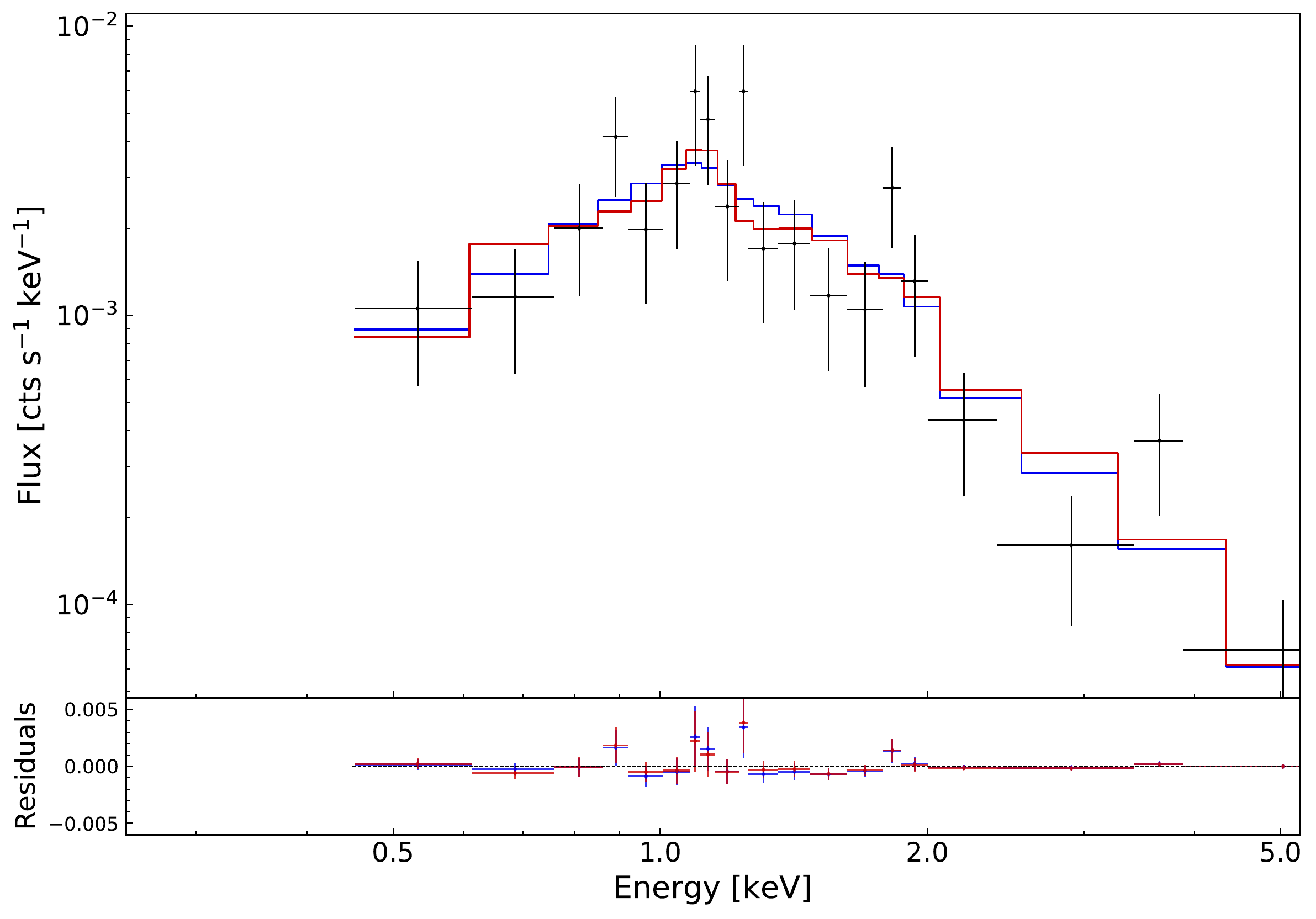}
\caption{\label{xspe} X-ray spectra of CU Vir binned to a minimum of five counts per bin; {\it XMM-Newton} pn/black\,+\,MOS/green (top) and {\it Chandra} ACIS-S (bottom) with thermal (red) and thermal+nonthermal (blue) model respectively, the residuals are shown in the sub-panels.}
\end{figure}

\begin{table*}[t]
\caption{\label{sres}Spectral fit results for CU Vir@79~pc, {\it XMM}: pn+MOS, {\it Chandra}: ACIS-S.}
\begin{center}
\begin{tabular}{lrrrr}\hline\hline\\[-3.1mm]
Par. & \multicolumn{3}{c}{Obs.} & unit\\\hline\\[-3mm]
& \multicolumn{2}{c}{XMM - 2011}& Chandra - 2017 &\\\hline\\[-3mm]
kT1 &0.12\,$^{+ 0.25}_{- 0.09}$  & 0.72\,$^{+ 0.12}_{- 0.15}$  & 0.23\,$^{+ 0.72}_{- 0.14}$ & keV  \\[1mm]
kT2 &0.76\,$^{+ 0.22}_{- 0.14}$  & 2.54\,$^{+ 1.30}_{- 0.60}$  & 2.65\,$^{+ 1.60}_{- 0.69}$ & keV  \\[1mm]
kT3 &2.62\,$^{+ 1.95}_{- 0.69}$  & --                        & -- & keV  \\[1mm]
EM1 & 0.23\,$^{+ 0.45}_{- 0.20}$ & 0.27\,$^{+ 0.09}_{- 0.08}$ &  0.50\,$^{+ 0.37}_{- 0.40}$ &  $10^{51}$cm$^{-3}$\\[1mm]
EM2 & 0.28\,$^{+ 0.09}_{- 0.08}$ & 1.13\,$^{+ 0.22}_{- 0.22}$ & 1.67\,$^{+ 0.37}_{- 0.32}$ &  $10^{51}$cm$^{-3}$\\[1mm]
EM3 & 1.04\,$^{+ 0.23}_{- 0.22}$ & --                        & --&  $10^{51}$cm$^{-3}$\\[1mm]\hline\\[-3mm]
C-Stat {\tiny(d.o.f.)} & 45.6 (79)& 49.7 (81)           & 19.5 (16)& \\[1mm]\hline\\[-2mm]
$L_{\rm X}$ {\tiny (0.2-10.0 / 0.5-5.0keV)} & 2.8 / 1.9 & 2.5 / 2.0  & -- / 2.6 &  $10^{28}$\,erg\,s$^{-1}$\\[1mm]\hline\\[-2mm]
kT &  \multicolumn{2}{c}{ 0.92\,$^{+ 0.11}_{- 0.20}$} & 1.0                    & keV  \\[1mm]
EM &   \multicolumn{2}{c}{0.33\,$^{+ 0.12}_{- 0.11}$} & 0.32\,$^{+ 0.3}_{- 0.3}$ &    $10^{51}$cm$^{-3}$\\[1mm]
$\alpha$ &  \multicolumn{2}{c}{2.0$^{+ 0.2}_{- 0.2}$} & 2.4\,$^{+ 0.4}_{- 0.5}$  & [PhoIndex]\\[1mm]
norm {\tiny (at 1 keV)}&  \multicolumn{2}{c}{4.5\,$^{+ 1.0}_{- 1.0}$} & 9.7\,$^{+ 2.3}_{- 2.8}$ & $10^{-6}$ph\,keV$^{-1}$\,cm$^{-2}$ \\[1mm]\hline\\[-3mm]
C-Stat{\tiny(d.o.f.)} &  \multicolumn{2}{c}{50.4 (81)} & 18.4 (17) & \\[0.5mm]\hline\\[-2mm]
$L_{\rm X}$ {\tiny (0.2-10.0 / 0.5-5.0keV)} & \multicolumn{2}{c}{2.9 / 1.9} & -- / 2.5 &  $10^{28}$\,erg\,s$^{-1}$\\[1mm]\hline\hline
\end{tabular}
\end{center}
\end{table*}

Given the absence of detectable strong variations in the X-ray brightness of CU~Vir, the spectra were extracted for the full observation each. For the 2011 observation we obtain two {\it XMM-Newton} spectra, one for the pn detector and one for the two MOS detectors combined; for 2017 the {\it Chandra} ACIS-S spectrum is used. 
The {\it XMM-Newton} and {\it Chandra} observation are modeled independently, each with a multi-temperature thermal model and a thermal+powerlaw model. Absorption is found to be negligible and is compatible with zero in all applied models. It is thus neglected in modelling, however due to the interdependence with the emission measure of the cooler plasma the X-ray spectra are not suited to derive meaningful constraints. Similarly, \cite{koch14} find no evidence for circumstellar material in the vicinity of CU~Vir in their analysis of hydrogen lines.
Abundances were set to solar values as likewise any potential deviation cannot be meaningfully explored with the data.
The X-ray spectra and respectively best fit models are shown in Fig.~\ref {xspe}.

For X-ray spectra with CCD type spectral resolution and moderate SNR, the two models look quite similar and not surprisingly they achieve a similar quality in fitting the spectra. To describe the {\it XMM-Newton} pn+MOS spectra we find a three temperature or a single temperature plus powerlaw model sufficient. The {\it Chandra} ACIS spectrum can be described with a two-temperature model or a single temperature plus powerlaw model, whereas the thermal component is poorly constrained and was fixed at 1~keV, i.e. roughly the {\it XMM-Newton} value. While several derived spectral model parameters are only moderately constrained, the general spectral trends are robust and independent of the used dataset or modelling approach.
A distance of 79~pc is adopted for CU~Vir to convert the X-ray fluxes to luminosities, for the applied models we find $F_{\rm X} = 3.8 \times 10^{-14}$~erg\,cm$^{-2}$\,s$^{-1}$ as mean value in the 0.2\,--\,10.0~keV band.
The X-ray luminosities used to study long term variability are calculated additionally in the energy range 0.5\,--\,5.0~keV, i.e. where fluxes are sufficiently well constrained by both datasets.
Our modelling results are summarized in Table~\ref{sres}.

Comparing the {\it XMM-Newton} observation in 2011 with the one from {\it Chandra} in 2017, the spectra are overall similar and the derived model parameters like plasma temperatures and/or power-law slope are, except for the normalization, within errors fully consistent between the two observations.

Independent if a purely thermal or thermal+powerlaw spectral model is used, the observed source flux is strongly dominated by the hottest plasma or the powerlaw component.
The thermal models are dominated by hot plasma at temperatures at 20\,--\,30~MK, cooler plasma at temperatures less than 10~MK is present but found to be a minor contributor at about 30\%. The 3-temperature modelling of the {\it XMM-Newton} data indicates the presence of plasma with a broad temperature distribution below 10~MK. While the detailed shape of the emission measure distribution (EMD) is poorly constrained, the full EMD is always dominated by hot plasma.
The average plasma temperatures $T_{\rm X}$ are in the 22\,--\,25~MK range in all thermal models. 

In thermal + nonthermal models, the emission is dominated by a powerlaw component with a slope in the range of 1.9\,--\,2.2 in both datasets. The modeled plasma temperature of the thermal component is about 10~MK and thus slightly higher than the average temperature derived for the cooler component(s) in the purely thermal model. Additional cooler plasma components might be present, but are at most a minor contributor. While the thermal component is poorly constrained in the {\it Chandra} data, a pure powerlaw model results for the {\it XMM-Newton} data in a very poor fit (C-Stat: 94.1/83) and can be virtually ruled out. As a cross-check we performed a 'goodness' Monte-Carlo test in Xspec on the {\it XMM-Newton} spectra binned to a minimum of 15~counts. In a run with 1000~simulations 96.80\,\% of the model realizations have a better test statistics than the data, confirming the low probability of the single powerlaw model.

Taking the results derived from the {\it XMM-Newton} observation, the powerlaw component contributes about 70\,\% of the flux in the 0.2\,--\,10.0~keV band and 95\,\% to the flux above 2.0~keV. Even more extreme is the {\it Chandra} observation, where we find a roughly 30\,\% higher X-ray luminosity (0.5\,--\,5.0 keV band) that is mostly caused by an increase of the emission measure in the hottest plasma component or the normalization of the powerlaw component that contributes here about 90\,\%. If the flux difference between 2011 and 2017 is associated to long-term variability or due to the fact that the observations have a different phase coverage remains open. Overall, the re-detection with {\it Chandra} at a comparable brightness of $L_{\rm X} = 2-3 \times 10^{28}$~erg\,s$^{-1}$ establishes CU~Vir as a persistent and hard stellar X-ray source.

\subsection{Activity ratios}

Using $L_{\rm bol} \approx 100~L_{\odot}$ \citep{koch14} we obtain $\log L_{\rm X}/L_{\rm bol} = -7.1$.
The activity ratio $\log  L_{\rm X}/L_{\rm bol}$ of CU~Vir is two orders higher than the value derived from the correlation obtained by \cite{naze14} for magnetic early/mid B stars, which predicts $\log  L_{\rm X}/L_{\rm bol} \approx -9.2$ from luminosity scaling. Adopting the estimated mass loss rate of $\dot{M} \approx 10^{-12}~M_{\odot}$\,yr$^{-1}$ from \cite{leto06}, one predicts again $\log  L_{\rm X}/L_{\rm bol} \approx -9.2$. Even considering the scatter of about one order of magnitude that is present within these relations, CU~Vir appears as a strong X-ray active star if compared to more massive stars powered solely by magnetic channelled wind shocks.

Using the average radio spectral luminosity of $L_{\rm rad} \approx 3 \times 10^{16}$ erg\,s$^{-1}$Hz$^{-1}$ \citep{leto06}, we obtain $\log  L_{\rm X}/L_{\rm rad} = 12.0$.
The derived X-ray/radio brightness ratio strongly violates the Guedel-Benz relation, that predicts $\log L_{\rm X}/L_{\rm R} = 15.5$ for coronal emission \citep{gue93}.
Even considering the scatter of about half an order of magnitude that is present within these relations, CU~Vir appears as an strong radio active star
if compared to coronal sources that are powered by magnetic activity.

\section{Discussion}
\label{dis}

The derived X-ray properties of CU~Vir and potential X-ray generating mechanisms are discussed in the following.

\subsection{Intrinsic vs. extrinsic emission}

The sharp {\it Chandra} PSF and good positional match make a chance alignment with an unrelated object very unlikely. This is specially true at its galactic latitude of +58.6~deg and its X-ray spectra without any hints on absorption on the line of sight and requiring multiple thermal or thermal+nonthermal components. In contrast, a very close and so far undiscovered companion to CU~Vir is not ruled out by our data. Given the measured X-ray luminosity of $\log L_{\rm X} = 28.4$, a late-type star is a viable option. However, the extraordinarily hard X-ray spectrum argues against this hypothesis for our target.

Even active M dwarfs emitting a several times $10^{28}$~erg\,s$^{-1}$ in X-rays, i.e. those that are comparable or moderately X-ray brighter than CU~Vir, have significantly cooler coronae with average temperatures of 6\,--\,8~MK and spectral energy distributions that peak around 8~MK \citep[e.g.][]{rob05}. Stars that possess coronae with average temperatures at quasi-quiescent level as observed for CU~Vir exist, however these have so far only been observed in stars with X-ray luminosities around or exceeding the $\log L_{\rm X} = 30$ level, i.e. objects that are hundred times X-ray brighter. These luminosities are ruled out, as the distance to CU~Vir and thereby flux to luminosity conversion, is well established. The recent Gaia DR2 gives $d= 72 \pm 2$~pc, even reducing the above luminosities by about 20\,\%.

The shape of the observed X-ray light curves, each obtained over several tens of ks, also clearly argues against an origin in a very strong flare that could produce sufficient amounts of correspondingly hot plasma. Furthermore, the repeated X-ray detection at a similar brightness and spectral hardness in observation separated by several years suggests that the derived X-ray properties reflect the typical state for CU~Vir.

Overall, coronal emission from a stellar companion might contribute, but is an unlikely explanation for the bulk of X-rays that are observed from CU~Vir. As the possibility of a chance alignment with an extragalactic or galactic counterpart is negligible, intrinsic mechanisms that are capable of generating the observed X-ray emission are considered to be the most likely explanation for the X-ray detection.

\subsection{CU Vir in context of magnetic intermediate mass stars}
The strong violation of both stellar scaling relations by CU~Vir highlights its outstanding character also at X-ray energies and in combination with its moderate X-ray luminosity and hard X-ray spectrum it is so far unique in the late-B/early-A star regime. Although several stars in the MCP sample presented in \cite{rob16} have similar characteristics in at least one or two of the relevant parameters, indicating similarities in their X-ray generating mechanism, none is as extreme as CU~Vir. 

Even typical magnetic early-B stars with similar $\log L_{\rm X}/L_{\rm bol}$ typically have significantly cooler/softer X-ray emission \citep{osk11}. Nevertheless, comparably hard X-ray spectra combined with high $\log L_{\rm X}/L_{\rm bol}$ are also known for a few magnetic early-B stars like HD~182180/HR~7355 and HD~142184/HR~5907 \citep{naze14,leto17, leto18} or $\sigma Ori$~E \citep{san04b}, however these have X-ray luminosities that are by factors of several tens or even hundreds above the one of CU~Vir.

Intriguingly, also the X-ray spectral properties are quite similar between CU Vir and the fast rotating Bp stars HR~7355 and HR~5907 \citep{leto17, leto18}. Albeit, the X-ray luminosity of these Bp stars is one to two orders of magnitude higher, the basic X-ray spectral properties and X-ray activity as well as X-ray/radio ratios are comparable. A discussion relating to the auroral model proposed for these stars is given in the next section. In Table~\ref{comp} we compare relevant parameters to highlight their similarities and differences.

\begin{table}[t]
\begin{center}
\caption{\label{comp}CU Vir vs. early Bp stars}
\begin{tabular}{lccc}\hline\\[-3mm]
 & CU Vir & HR 5907 & HR 7355\\\hline\\[-3mm]
$M_{*}$ [M$_{\odot}$] & 3.1 & 5.5 & 6.0 \\
$P_{\rm rot}$ [d] & 0.52 & 0.51 & 0.52\\
$B_{\rm p}$ [kG] & 3.8 & 15.7 & 11.6\\\\[-3mm]

$\log L_{\rm X}$ [erg\,s$^{-1}$] & 28.4 &30.1 & 30.0\\
kT/$\alpha$ & 0.9/2.0 & 1.0/1.6 & 1.0/1.7 \\
$\log L_{\rm X}/L_{\rm bol}$ & -7.1 & -6.4 & -6.5\\
log $L_{\rm X}/L_{\rm rad}$ & 12.0 & 11.8 & 12.0\\\hline
\end{tabular}
\tablefoot{CU Vir (this work), HR 5907 \citep{leto18}, HR 7355 \citep{leto17}.}
\end{center}
\end{table}

\subsection{Purely thermal vs. thermal plus non-thermal X-rays}

In Ap/Bp stars the stellar magnetosphere offers several mechanisms capable of producing X-ray emission. The strongly magnetic star CU Vir possesses a primarily dipolar-like non-axisymmetric magnetic structure. Thermal X-ray plasma will be naturally created in wind shocks via the MCWS mechanism, but the weaker winds in late-B\,/\,early-A stars with terminal velocities of $V_{\inf}\approx 600$~km\,s$^{-1}$ are expected to create plasma with post-shock temperatures of a few up to about 10~MK \citep[e.g.][]{dou14}.
The 'classical' MCWS plasma is thus a suitable candidate for the cooler plasma component(s) in our spectral models, but insufficient to explain the hot, dominant component around 30~MK.

Overall, the main characteristics of CU~Vir derived here, i.e. faint but hard X-ray emission, are virtually the exact opposite of the original motivation for the development of the MCWS model by \cite{bab97} to explain the ROSAT data of the A0p star IQ~Aur, i.e. bright but soft X-ray emission.

Clearly, a mechanism that, depending on the adopted model, is producing the hottest plasma or non-thermal component is needed. Furthermore, this component is not a small extra, but actually the dominant contribution to the observed X-ray emission; comparing results from 2011 ({\it XMM-Newton}) with those from 2017 ({\it Chandra}) it is likely also the more variable component. 

The very hot plasma could arise from activity-like phenomena associated with magnetic spots or plasma captured in the magnetosphere, where in addition re-connection events might occur in the disk-like structure e.g. during break-out or infall events. 
However, compared to similar Ap/Bp stars like the well studied IQ~Aur \citep{rob11}, the required plasma temperatures of CU~Vir are extreme. We find quasi-quiescent X-rays for IQ~Aur with $\log L_{\rm X} \approx 29.6$~erg\,s$^{-1}$ and $T_{\rm X} \approx 8$~MK and for CU~Vir with $\log L_{\rm X} \approx 28.4$~erg\,s$^{-1}$ and $T_{\rm X} \approx 25$~MK.
Similarly hard X-ray spectra are only seen during a flare event in IQ~Aur, which is associated with very hot thermal plasma at several tens of MK as deduced from the detection of a strong 6.7~keV \ion{Fe}{XXV} emission line complex. However, in IQ~Aur this is a transient phenomenon, whereas the very hot component is clearly associated with the quasi-quiescent state of CU~Vir.

An alternative scenario, involving a non-thermal component that describes auroral X-ray emission, was proposed by \cite{leto17}, put forward to explain the X-ray emission from fast rotating magnetic early B star HR~7355.
It suggests, that besides the thermal emission originating from the MCWS that heats plasma up to a few MK, non-thermal emission should be present in fast rotating magnetic Ap/Bp stars.
In this scenario the X-ray emission originates from the non-thermal electrons upon impact on the stellar surface. The electron population would be identical to the one, that is also responsible for the gyrosynchrotron radio emission. 
The non-thermal electron population able to produce the incoherent gyrosynchrotron emission of CU~Vir
has a power-law energy distribution \citep{leto06} with a low energy cutoff of $\approx 100$ keV \citep{tri04}.
Such non-thermal electrons, accelerated by magnetic reconnection events occurring in the current sheets regions located far from the star about 15 stellar radii, precipitate toward the stellar surface.
The energy budget of these precipitating electrons is compatible with what observed during the Sun flares, where also hard X-ray emission is detected at the footprints of
the magnetic loops \citep[see][and references therein]{asch02}.
In this case the hard X-ray component would be a power-law component generated by thick target bremsstrahlung emission from the non-thermal electrons. 
Indeed, CU~Vir could be understood as a down-scaled version of HR~7355, whose radio luminosity as well as its X-ray luminosity is about a factor 30 higher.
What causes these phenomena in CU~Vir and similar objects remains so far open, but fast rotation and strong dipolar fields have been identified as common attributes.
Further hard X-ray observations at photon energies higher than 5~keV could be very useful to definitively confirm this scenario.

The X-ray spectrum of CU~Vir alone is not able to definitely support the scenario where the hard X-ray component has auroral origin.
The thermal and the non-thermal scenario are both viable and even both may be at work simultaneously. Furthermore, the thermal component in itself may be a composite that is partly MCWS, magnetospheric activity or reprocessed auroral emission in nature. However, in any case some extraordinary X-ray generation mechanism has to be present in any intrinsic emission scenario.

\section{Summary}
\label{sum}

We detect persistent X-ray emission from the A0p stellar radio pulsar CU Vir. Its main characteristics can be summarized as follows.

\begin{enumerate}
\item The X-ray luminosity is with $L_{\rm X} \approx 3 \times 10^{28}$~erg\,s$^{-1}$ moderate, light curves show only minor variability.
\item The X-ray spectra are very hard, their modelling requires multi-thermal models dominated by hot plasma at 25~MK or thermal + non-thermal plasma components.
\item The X-ray emission is very likely intrinsic; modified MCWS models e.g. including aurorae are promising, but details of its generating mechanism are open.
\item The {\it XMM-Newton} (2011) and {\it Chandra} (2017) data give similar X-ray properties, indicating that these are quite stable and persistent.
\item The A0p type star CU~Vir is virtually a down-scaled version of the early-B type MCP stars characterized by a similar X-ray spectrum and X-ray/radio luminosity ratio. 
\end{enumerate}

The X-ray and radio features observed in CU~Vir can be overall explained by the scenario proposed in \cite{leto17}; it thus holds both in the early-B as well as in late-B\,/\,early-A fast rotating MCP stars. As a future outlook we plan a combined X-ray and radio data modeling and to extend the simulations of the radio emission of CU~Vir up to the millimeter wavelengths.

\begin{acknowledgements}
This work is based on observations obtained with {\it Chandra} and {\it XMM-Newton}.
J.R. and LMO acknowledge support from the DLR under grant 50QR1605 (JR) and FKZ50OR1508 (LMO). LMO acknowledges partial support by the Russian Government Program of Competitive Growth of Kazan Federal University.

\end{acknowledgements}

\bibliographystyle{aa}
\bibliography{cuvirXray.bbl}

\end{document}